# Direct Charge Ejection for Chemical Electric Generation

Anthony Zuppero[a], Thomas J. Dolan[b]

[a] neofuel.com, 2631 Buena Tierra Rd, Pollock Pines, CA, 95726, USA
[b] NPRE Department, University of Illinois, 104 S. Wright St., Urbana, IL, 61801, USA

## Abstract

A particular type of surface chemical reaction resulted in direct charge ejection. Several different experimental observations strongly support the hypothesis that nearly all of the molecular *kinetic* energy of a highly excited vibrating molecule was transferred to the *kinetic* energy of a single electron in a conducting lattice, leaving the molecule in nearly its ground state. This kinetic energy transfer represents the counterpart to the potential energy transfer in a battery, where electrochemical *potential* energy of reactants is transferred to the electrical *potential* energy of a separated charge. Here chemical reaction energy is transferred directly to the kinetic energy of a charge instead of to the potential energy of a charge. These observations indicate a breakdown of the Born Oppenheimer approximation.

## Introduction

Direct charge ejection became potentially useful when recent experimental evidence showed that a substantial fraction of the kinetic energy of a molecular vibration seemed to be converted into the kinetic energy of a single charge, and for the situation where the molecular vibration was energized directly using a chemical reaction. The stages of these observations are summarized in Figure 1.

## Experimental Observations

In the first observations, electron emissions associated with energetic oxygen or halogen gas molecules interacting directly with alkali metal surface were measured as extremely tiny currents in a vacuum tube diode. The fraction of electrons per reaction was apparently many orders of magnitude less than 1. An energetic reaction was needed to provide enough energy for the emitted charge (electron) to overcome ~ 2 eV metal / vacuum work function in order to leave the metal cathode and be collected at the vacuum tube collector plate. The process is referred to as "exoelectrons". Figure 1 with label "exo-electron" shows the configuration. [Greber-1997, Böttcher-1990, Gesell-1970, Hellberg-1995 ]

---

[a] patents@neofuel.com
[b] dolantj@illinois.edu

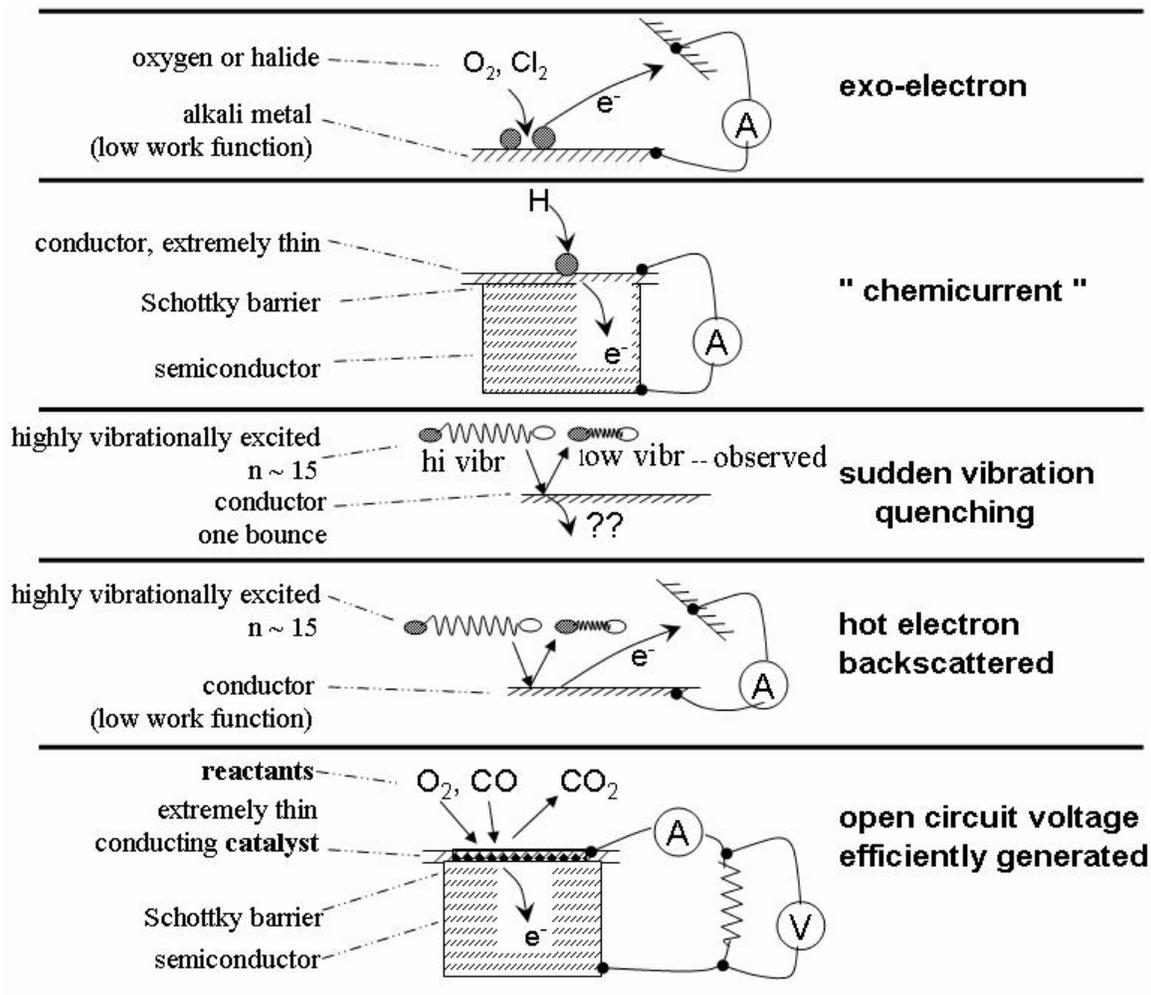

**Figure 1. Sequence of experiments showing chemical association reactions energizing direct charge ejection.**

Nienhaus and his team used a semiconductor instead of a vacuum as the charge collector, as shown in Figure 1 with label "chemicurrent". [Nienhaus-1999]

Glass and Nienhaus observed the counterpart, a hole chemicurrent in Mg surface oxidation. [Glass-2004, Nienhaus-2006] The observation of the hole counterpart shows that the charges involved are quasi-particles. Both electrons and holes in metals, semiconductors and insulators are quasi-particles.

In both the exoelectron and chemicurrent cases the hot electron generated by a chemical gas-surface reaction traveled against an electrical potential to reach the collector. However, the value of the work function-equivalent at the metal / semiconductor junction was only of order 0.5 eV, and was far lower than the metal / vacuum work function, of order 2 eV. As a result, Nienhaus et al. could collect those electrons with energies less than the exo-electron energy ~ 2 eV ( but > 0.5 eV).

The possibility to use a lower electron reaction energy expanded the set of reactants to include atomic hydrogen, atomic oxygen and others, adsorbing onto clean, oxide free, semi-noble metal surfaces (a surface-substrate reaction). Cryogenic temperatures were required because of the tiny currents. All these reactions can be classified as "association reactions" because a gas atom or molecule forms a chemical bond with the surface, even though the bond may be weaker than a molecular bond.

Zuppero et al applied for a patent claiming that vibrational energy alone could energize electrons in a configuration where a metal-semiconductor diode would convert electron kinetic energy into useful potential. [Zuppero-2001]

Both Nienhaus and Zuppero emphasized the need for extremely thin cathode conductor thicknesses, less than the mean free path of a ballistic electron with energy in the range ~ 0.5 - 2 eV. The electron (or hole) must not be stopped before it leaves on a path to the semiconductor electron collector. This boundary condition of "extremely thin" posed a severe experimental problem because such mean free paths are typically in the range 2 - 20 nm. When the required metal layers are this thin the deposition and control of the metal layer become difficult. The metal layers tend to agglomerate into clusters on a ceramic, such as a semiconductor surface. Agglomeration destroys the "continuous surface" characteristic and can destroy the diode properties.

Huang et al. observed what appeared to be prompt (tens of femtosecond) transfer of the highly excited molecular vibration energy to a highly conducting surface, in a single, gas-surface encounter. An association reaction begins with an initial state like that of a highly vibrationally excited molecule at the moment when its constituents are most stretched. [Huang-2000]

Huang et al. suggested electronic involvement. Using a laser they energized a molecule to a single, highly vibrationally-excited state (~1.5 eV), which was a substantial fraction of the bond energy. They observed that much of the vibrational energy was suddenly quenched when the molecule bounced off a highly conducting, noble metal substrate. They argued that the energy must have been transferred to the highly conducting substrate during a single collision. They deduced that the energy must have gone entirely to a single electron, which would have instantaneously received ~ 1.5 eV. The experimenters observed essentially no quenching on an insulator. Figure 1 with label "sudden vibration quenching" shows this experiment.

Huang et al. referred to the process as "electron jump" because they hypothesized that a low energy, conductor electron would "jump" from the conductor onto the highly vibrating molecule when the molecule was in it's maximum stretched state. In that state, the electron affinity of the molecule is most energetically favorable to accept an electron from somewhere. Half a vibration period later, with the molecule still close enough to the conductor for electrons to jump to and from the conductor, the same molecule would be in its maximum compressed state. At maximum molecular compression the electron affinity is least favored, and the affinity is most favorable for an electron to be ejected from the molecule.

Huang et al. argued that the entire process time of a molecule / surface collision (~ 100 femtoseconds) is orders of magnitude too short to permit vibrational energy transfer. This left only electron energy transfer, and, in principle, energy transfer to only one or a few electrons.

White et al. used low work-function conductor surfaces with the same NO molecule and experiment of Huang et al, and measured energetic electrons backscattered out of the metal surface. The backscattered electron energy supported the single-electron hypothesis of Huang et al. The experiment demonstrated that the fraction of hot electrons produced was greater than $10^{-3}$ and could be almost 1, depending on assumptions. Figure 1 labeled "hot electron backscattered" illustrates this process. [White-2006]

With guidance and funding from NeoKismet LLC, Ji et al. used the surface catalytic reactions of adsorbed CO with adsorbed O to energize a highly vibrationally-excited initial state of $CO_2$, and collected the electron kinetic energy using the Schottky diode formed by a thin, catalyst conductor ( Pt or Pd) with an n-type semiconductor. [Ji-2005a, Park-2006a, Park-2006b]

In one set of experiments Ji et al. oxidized CO on a diode formed by a nanometers thin Pt catalytic surface and an n-type, $TiO_2$ semiconductor. Ji measured approximately 3 electrons surmounting the ~ 0.5 eV Schottky barrier of each 4 reactions on a Pt catalyst surface. Reaction temperatures up to 200 Celsius were used at atmospheric pressure, sharply contrasting the ultrahigh vacuum and cryogenic temperatures of prior experiments. [Ji et_al 2005b]

In another similar, most curious and telling experiment, room temperature CO oxidation on a diode with a Pd catalytic surface confirmed the dominant hot electron production and demonstrated an "open circuit voltage." Figure 1 labeled "open circuit voltage efficiently generated" shows this experimental setup. [Ji et al.-2005a,b,c] This experiment used a 5 nm Pd nanolayer on a 2 mm diameter, n-type $TiO_2$ semiconductor to form a Schottky diode. This Pd/$TiO_2$ diode sustained 0.9 micro-Amps/$cm^2$ and generated an open-circuit voltage of 0.68 V. This current (Figure 2) corresponds to about 0.0038 reactions per site per second. The 0.0038 reactions per site per second at room temperature are completely consistent with known reaction rates at room temperature. Ji et al. used $TiO_2$ as the semiconductor because McFarland had demonstrated 0.5 eV across an Au/$TiO_2$ diode with a mere 15 micro-Amps /$cm^2$. [McFarland-2003]

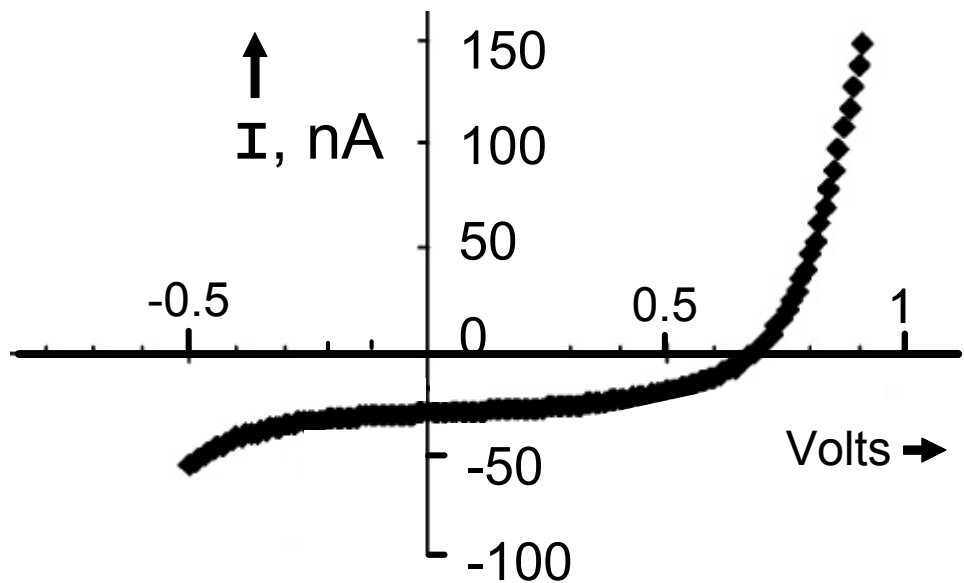

**Figure 2.** Volt-Ampere characteristics of Pd/TiO$_2$ direct ejection electric generation under 40 Torr of CO and 100 Torr of O$_2$ (at 1 atm, He synthetic air) at 293 K, with I$_{sc}$ = 29.1 nA and V$_{oc}$ 0.68 Volts. [From Ji-2005c . ]

This direct charge ejection could become useful as an electric generator if the chemical association reaction rate on a catalyst surface could be accelerated to be greater than the equivalent of about 500 reactions per catalyst surface site per second at room temperature. However, these conditions have not yet been met.

This sequence of observations provides a basis to conclude that, under some conditions, molecules in highly vibrationally excited states can transfer a substantial fraction their kinetic energy to a small number of charges, if the excited molecules are in atomic-scale contact with a substrate containing a high density of unoccupied energy states comparable to the energies associated with the charge motion.

In order to avoid confusion we will use the following terminology:
- Motion of an electron from a molecule into a substrate is called "ejection".
- Motion of an electron from a substrate into a molecule is called "jump".
- The generic name for these process is called "charge transfer".

## Association Reaction As Initializing Agent

For practical applications a key element is to promote charge transfer directly from the energetic reactants. An association reaction on or in a conductor provides such a stimulus. For example, consider the association of adsorbed CO with adsorbed O on a catalyst conductor surface. Each adsorbed reactant is initially loosely bonded chemically with the catalyst surface and each is in its own potential well. When an adsorbed CO in one potential well is next to an adsorbed O in an adjacent potential well, they might never react because they are too far apart and separated by an "activation barrier," which is the potential wall between them. However, thermal motions or tunneling can permit the reactants to interact, in spite of their trapping by the substrate. Their separation is similar to that of a highly stretched molecule formed by the association of the adsorbed CO with adsorbed O. Such a highly stretched molecule is almost at its breaking point, in an initial condition of a highly vibrationally excited molecule.

The potential distribution is shown in Figure 3.

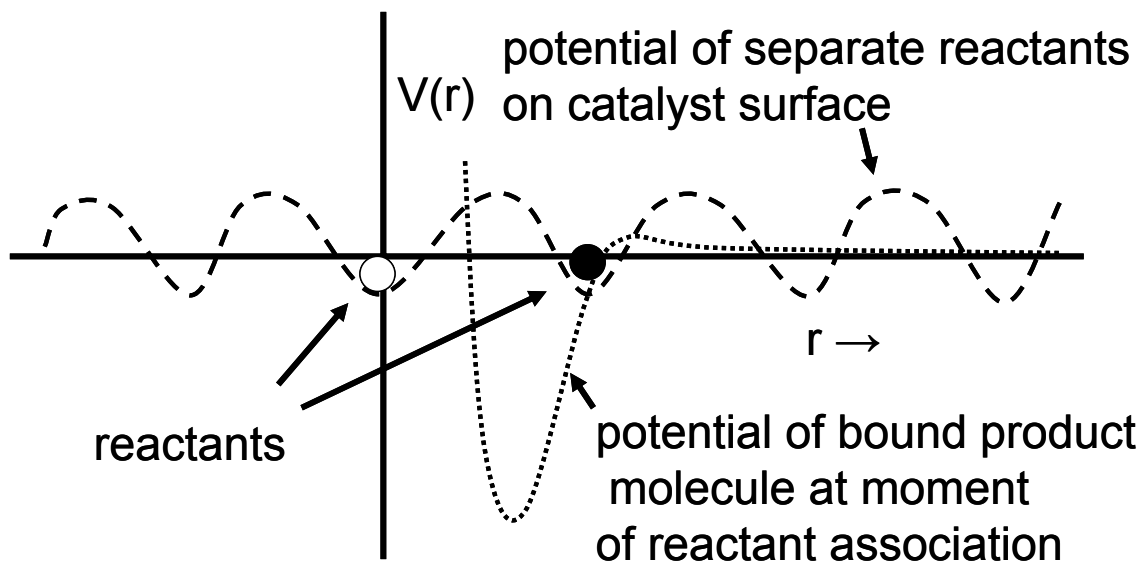

**Figure 3. Reactants in separate potential wells on catalyst surface interact to form a product in a state like that of a most-stretched molecule, an initial state for a highly vibrationally excited molecule.**

This initialization then creates a molecule directly on the surface with a high affinity for the charges (electrons) of the substrate. This is identical to the highly vibrationally excited molecule of Huang 2000 at the moment it contacts the conducting surface.

An energy level diagram for this surface association reaction of adsorbed CO with adsorbed O on a Pt surface (Figure 4) suggests $E_2 + H_2 = 1.46$ eV of energy is available from the top of the potential well (which keeps the reactants apart) to the bottom of the potential for the associated molecule (product molecule ground state). [Kwong-1988]

Figure 5 shows a similar diagram for a Ru catalyst. [Stampfl-1999]

A reaction on a catalyst surface helps channel the chemical energy of two reactants efficiently into highly vibrationally excited states. The surface denies rotational and translational dilution of the reaction energy and provides a reaction channel along a line between reactants.

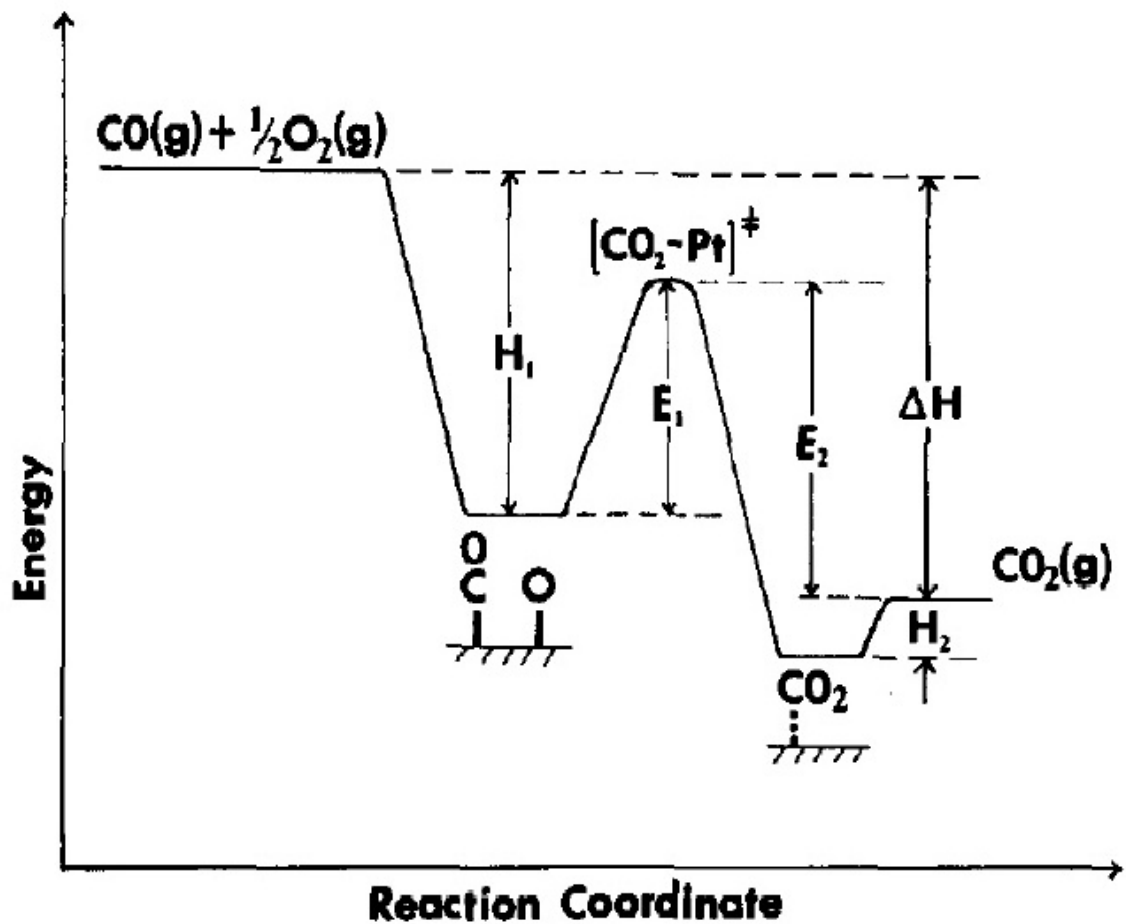

**Figure 4, energy level diagram for surface association reaction of adsorbed CO with adsorbed O on a Pt surface, copied directly from Kwong-1988. The potential energy diagram of CO oxidation on Pt(111) surface at low CO and O coverage. ΔH = 2.93 eV /mol. $H_1$ ~2.68 eV/mol, $H_2$~ 0.21 eV /mol, $E_1$ ~ 1.04 eV /mol, $E_2$~ 1.25 /mol. maximum stretch to maximum compression ~ 1.46 eV / mol. [from Kwong-1988]**

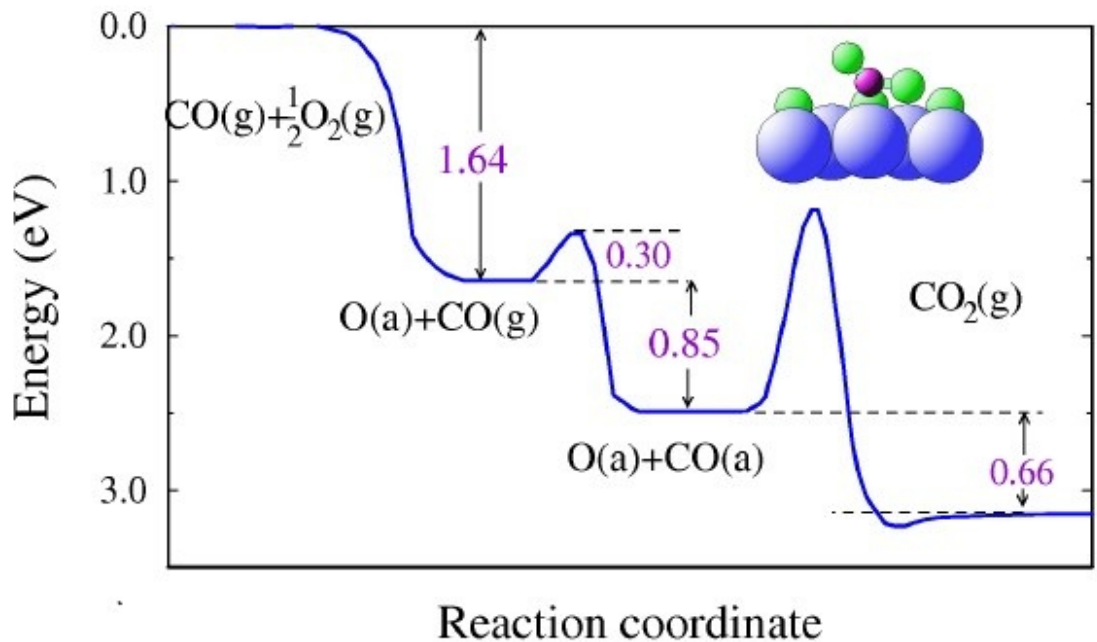

**Figure 5, Energy level diagram calculated for the Langmuir-Hinshelwood mechanism for surface oxidation of adsorbed CO with adsorbed O on Ru(001), showing approximately 1.8 eV available. The large, small, and small darker circles represent Ru, O, and C atoms, respectively. [From Stampfl-1999]**

## Proposed Explanation of Direct Charge Ejection

Auerbach ( coauthor of Huang-2000) used the electron affinity to help explain and quantify the electron ejection. Figure 6 shows that a charged, highly vibrationally-excited molecule has a potential energy minimum at a slightly larger separation of N and O than the same system with the electron ejected. Observing the overlay of the potential energy vs. the N-O separation distance for the charged and uncharged species, we note that at the inner turning point the highly vibrationally-excited N-O$^-$ can transition to the potential of the uncharged N-O if the energy is transferred to an electron. The conducting substrate provides a high density of such electron states for which a nearly-resonant transfer can occur. An insulating substrate does not.

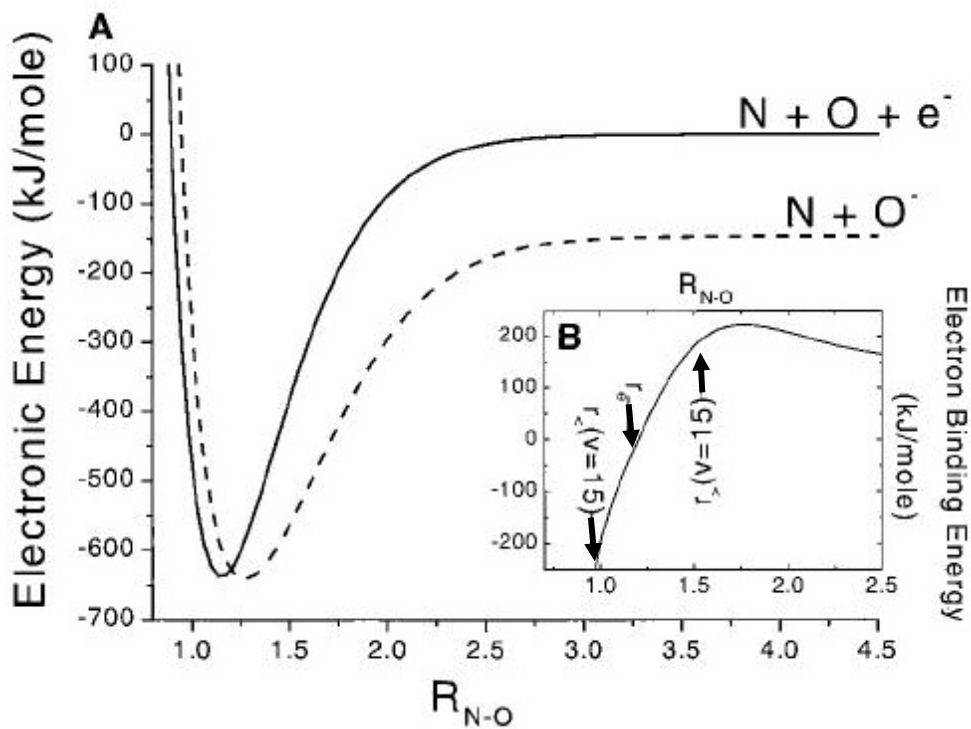

**Figure 6. Electron energy vs. N-O separation to elicit electron affinity. (A) ab initio calculation of NO and NO⁻ showing energetic constraints on electron transfers. (B) The difference between the two potential curves for a case with ν = 15. The arrows indicate the most compressed, equilibrium and most stretched points of classical N-O molecule. [From Fig.1 of Huang-2000]**

According to Auerbach, the electronic energy diagrams show charge-phobic and charge-philic regions of a molecule during its vibration phases. [Huang-2000]
The molecule is most charge-phobic when most compressed, favoring charge ejection, and most charge-philic when most stretched, favoring electron jump from the substrate to the molecule. Silva shows this in more detail. [Silva-2001]

The system thus comprises a molecule that is initially in a highly vibrationally-excited state and a substrate containing a high density of empty, charged quasiparticle states (electrons or holes) with energies near that of the vibrational state. The system can transition to a state comprising a molecule in a lower vibration state and a substrate with one of the many charged quasiparticle states filled.

Direct ejection of a high energy charge is highly favored during the compressed phase of the molecular entity vibration because the charge-donating, vibrating molecule is charge-phobic, and because the charge-receiving substrate has a large number of accepting states.

Electron jump of a lower energy charge from the substrate to a molecule is favored in the Huang-2000 experiment during the most-stretched part of the vibration because there are a large number of filled, available, lower energy electron states in the conductor substrate and because the most-stretched molecule is highly charge-philic. There is no charge jump without a substrate containing a high density of charged quasiparticles that can "jump" onto the molecule. The Huang experiment demonstrated that when their highly vibrationally-excited molecule contacted a highly insulating LiF surface no detectable vibrational energy transfer occurred.

The counterpart, charge transfer involving a hole, has also been observed. [Glass-2004 and Nienhaus et al-2006]

We suggest that this combination of occupied vibrational and empty electronic energetic states in a polyatomic system is similar to that where the Quantum Defect Theories (QDT) apply. Direct charge transfer will then correspond to QDT autoionization, and the initial reactants correspond to QDT Rydberg states. Pratt's review provides a theoretical basis to support the use of Quantum Defect Theory to quantify the observations. [Pratt-2005]

## Conclusion

Direct charge transfer appears to result when a highly vibrationally excited molecule is in direct contact with a conducting substrate. The charge can carry away a substantial amount ($> 0.5$ eV) of the vibrational energy, if a substrate, such as a conductor, with a high density of empty, kinetic energy states is available for the transferred charge.

Theoretical analysis involves a breakdown of the Born-Oppenheimer approximation, but is not yet completely understood. The observations involve:

- The highly vibrationally energized molecules being in direct contact with a substrate with high density of unfilled, matching energy states.
- Most of the vibrational kinetic energy being transferred to the kinetic energy of single electron or hole in the substrate, which is one of the unfilled, matching energy states.

The system undergoes a prompt transition to a final state where
- the molelcule is in a lower vibrational state and
- one of the many quasiparticle states near the energy difference is occupied.

The process is strikingly similar to the kind of autoinozation described by QDT.

Association reactions in or on a conductor can energize the extreme molecular vibrations and lead quickly to direct charge ejection or electron jump. A substantial fraction of the association reaction energy can be concentrated in a few electrons or holes. Such a situation can occur when chemical reactants, such as fuel and oxidizer, react on a catalyst surface. Such chemical reactions embedded inside a conductor can also produce similar direct charge transfer.

-------------